# Performance Analysis and Optimization of Dowels in Jointed Concrete Floors

**Ghauch Z.[1], Karam G.[2]**

[1]Undergraduate student, Department of Civil Engineering, Lebanese American University, Byblos, Lebanon, e-mail: zdghaouche@gmail.com

[2]Associate Professor, Department of Civil Engineering, Lebanese American University, Byblos, Lebanon, e-mail: gkaram@lau.edu.lb

## Abstract

This study examines the performance of traditional round dowels in concrete floors and attempts to optimize the shape of dowel bars through Finite Element (FE) analysis. A new type of Double-Tapered Round (DTR) dowels is proposed, and the performance of DTR dowels is compared to that of traditional cylindrical dowels. Linear Elastic (LE) analysis is performed in Abaqus (v-6.11) in order to identify the optimized geometry of DTR dowels that would achieve (1) highest load transfer across adjacent slabs through shear action, and (2) lowest bearing stresses on the concrete. LE analyses are complemented by nonlinear FE analysis. The Riks method available in Abaqus/Standard, coupled with the Concrete Damaged Plasticity (CDP) model is used to simulate the degradation of concrete surrounding both DTR and traditional cylindrical dowels. Results obtained show that the use of DTR dowels can reduce bearing stresses at the face of the joint by as much as 2.2 times as compared to traditional cylindrical dowels. While adequate load-transfer is a crucial part for the proper performance of pavement structures, the load-transfer capacity of DTR dowels was found to be more effective over cylindrical dowels by as far as 116%. In the inelastic range, even after significant concrete degradation and steel yielding, DTR dowels maintained a higher load-transfer capacity than traditional cylindrical dowels, and also presented lower amounts of differential deflections across concrete floors. Finally, damage in the concrete matrix below the dowel was relatively more confined for the case of DTR dowels, as compared to traditional cylindrical dowels.

**Keywords:** Concrete pavements; dowels; finite element method.

## Introduction

**Background**

Despite the overall satisfactory performance of cylindrical steel dowels in concrete pavements, several flaws in the performance of cylindrical dowels have been identified. Among other things, traditional steel dowels are subject to: (1) corrosion of steel material, (2) restraint stresses and slab cracking associated with dowel misalignment, and (3) reduced load transfer efficiency due to dowel looseness and formation of voids in the concrete matrix surrounding the dowel. Dowel looseness is mainly associated with high bearing stresses exerted on the concrete due to the relatively low bearing area of cylindrical dowels (ACPA 2010).

We recall that the general equation for the deflection of a dowel structure extending infinitely into an elastic mass can be written as follows (Friberg 1938):

$$y(x) = \frac{e^{-\beta x}}{2\beta^3 E_s I}\left[P\cos\beta x + \beta\frac{Pz}{2}(\cos\beta x - \sin\beta x)\right] \qquad [1]$$

Where β is the relative stiffness of the dowel structure and is defined as follows:

$$\beta = \sqrt[4]{\frac{K\phi}{4E_s I}} \qquad [2]$$

In equation [2], $E_s$ and I are the modulus of elasticity and moment of inertia of the dowel bar, respectively, ø is the diameter of dowel bar, and K is the modulus of dowel-support, representing the pressure intensity (MPa) required to induce a 1mm settlement, and P is the downward shear acting on the dowel, and transferred from one PCC slab to the next. The bearing stress can be expressed as a function of the dowel deflection at the face of the joint ($y_o$) and the modulus of dowel support K as follows:

$$\sigma_o = Ky_o = \frac{KP}{2\beta^3 E_s I}\left[1 + \frac{\beta z}{2}\right] \qquad [3]$$

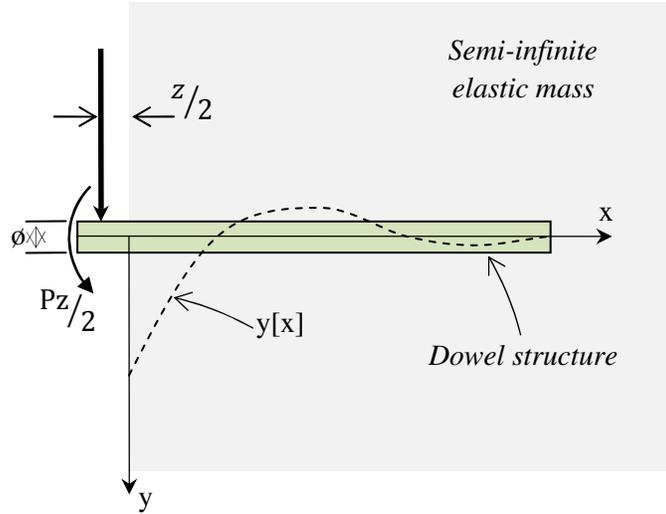

Figure 1 – Deflection shape of embedded dowel in semi-infinite elastic mass

Although the modulus of dowel support is of crucial importance in Friberg (1938) design equations, no appropriate theoretical method exists for the determination of this parameter. The only way for computing the modulus of dowel support is through experimental testing. Moreover, there is considerable uncertainty in design as to the selection of a proper value for the modulus of dowel support, since values of K typically range from 81.5 to 409 N/mm$^{-3}$.

Mannava et al. (1999) conducted a series of experimental load-deflection tests on cylindrical dowels with varying diameter, concrete strength, and joint width. The laboratory setup consisted of a single dowel bar connecting a loaded block to a reacting block; a static load applied through a hydraulic jack, and the resulting deflection profile along the embedded dowel was measured using a series of Linear Variable Displacement Transducers (LVDTs). The study observed that in order to match experimental deflections measured along the dowel with Friberg's analytical solution, several K-values were required. This raises the question as to how far the original assumption of uniform modulus of dowel support along the dowel is accurate.

Using the dowel deflection equation (equation [1]) presented by Timoshenko (1925) and revised by Friberg (1940), the shear in the dowel structure at any point x can be defined as follows:

$$V(x) = -e^{-\beta x}[(-\beta Pz - P)\sin \beta x + P \cos \beta x\,] \qquad [4]$$

Traditional methods for dowel design are based on empirical methods. The dowel diameter is directly selected based on the pavement thickness. An early empirical attempt for dowel design was to select the diameter of cylindrical dowels equal to one-eighth of the concrete slab thickness (PCA 1975). Further suggestions were made in terms of using a minimum dowel diameter of 1.25 in. in order to control the amount of faulting associated with bearing stress in the concrete. In 1991, PCA suggested using 1.25 in. dowel diameters for concrete slabs less than 10 in. thickness, and 1.5 in. dowels for slabs with higher thicknesses.

Another method for sizing dowels is based on the bearing stress in the concrete. The diameter of the dowel is selected so that the bearing stress exerted by the dowel on the concrete, as expressed by equation [3], remains below the allowable bearing stress of concrete given as follows (ACI 1956):

$$\sigma_{o,all} = \left(\frac{4-\phi}{3}\right) f_c' \qquad [5]$$

**Review of Commonly Used Dowel Shapes in Concrete Floors**

Traditional round dowels restraint the lateral movement of the slab parallel to the joint, movements that are associated with temperature variations. Dowel misalignment further restraint lateral displacements, thus increasing stresses and the potential for cracking in concrete floors. Schrader (1991) suggested a solution to the above problem by adopting square dowels equipped with compressible material attached along the vertical sides. In fact, the use of compressible material allows for unrestricted lateral displacement between adjacent slabs, thus eliminating the problem of restraint stresses and dowel misalignment. A high-density plastic clip, which significantly increased the cost of using square dowels, was later added in order to protect and hold the compressible material during construction.

Porter et al. (2001) investigated the performance of elliptical dowel bars in concrete slabs. The main advantage of elliptical dowels over traditional round dowels is in the increased bearing area and the resulting reduction in bearing stress on the concrete. However, the advantages of using elliptical dowel shapes are counterbalanced by difficulties in placement as well as availability concerns.

Walker & Holland (1998) proposed the use of rectangular and diamond plate dowels as an alternative to traditional round dowel bars. According to their study, diamond dowels constitute the optimal shape a plate dowel can take. As compared to round dowels, diamond plate dowels provide more material near the joint where the flexural, shear, and bearing stresses are high, and less material away from the joint where stresses decrease, while also allowing for horizontal movements of the slab parallel to the joint.

Further improvements in dowel shapes led to the development of tapered and double-tapered plate dowels. The main advantage in using such dowels, in addition to the relatively increased bearing area, is the fact that the dowel geometry by itself (void space around tapered sides) reduces or eliminates lateral slab restraint, without the need for compressible material attached at the vertical sides. Table [1] summarizes the commonly used dowel bars and plates in concrete floors.

Table 1 – Summary of commonly used dowel shapes in concrete floors [Walker & Holland (1998), ACPA (2010)]

| Dowel Bar Type | Applicable | | | Unrestrained Lateral Slab Movement | Increased Bearing Area | Other Advantage(s) or *Inconvenient (s)* |
|---|---|---|---|---|---|---|
| | Isolation joints | Contraction joints | Construction joints | | | |
| Round Dowel | ✓ | ✓ | ✓ | | | o Can be rotated, making it easy to remove forms |
| Elliptical Dowel | ✓ | ✓ | ✓ | | ✓ | o *Availability concerns and difficult placement* <br> o *Flexural bending around horizontal weak axis* |
| Square Dowel | ✓ | ✓ | ✓ | ✓* | | o *Increased cost of doweling due to plastic clips* |
| Rectangular Plate Dowel | | ✓ | ✓ | ✓* | ✓ | o *Increased cost of doweling due to plastic clips* |
| Diamond Plate Dowel | | | ✓ | ✓ | ✓ | o Optimized material use <br> o Cannot be misaligned |
| Tapered Plate Dowel | | ✓ | | ✓ | ✓ | o Optimized material use <br> o Cannot be misaligned |
| Double-Tapered Plate Dowel | | ✓ | | ✓ | ✓ | o Optimized material use <br> o Cannot be misaligned |

*Compressible material attached to the vertical faces allows for lateral slab movements

## Optimized Design of Dowel Bars

### Double-Tapered Round (DTR) Dowels

The design of cylindrical dowel bars can be significantly optimized by adopting more refined dowel shapes that would achieve (1) better load transfer capacity through vertical shear action, and (2) lower bearing stress exerted on the concrete seating. In this context, we investigate in this study a new type of Double-Tapered Round (DTR) dowel bars. The performance of tapered round dowels bars would have mainly three advantages over the commonly used dowel bars: (1) more optimized dowel shape that follows the shear distribution along the dowel, hence allowing for better load transfer, (2) higher bearing area in the region where the concrete is mostly stressed, and (3) as in the case of plate dowels, this tapered round dowel bar allows for lateral movement of the dowel due to the presence of a void space that allows for temperature shrinkage and contraction, hence eliminating any restraint stresses resulting from dowel misalignment.

Let us consider a dowel bar with geometry as shown in figure [2]. We assume that the diameter of the dowel ends is only a fraction α of the maximum diameter $\emptyset_0$ in the joint section. In the joint of width z, the dowel has a constant diameter $\emptyset_0$, while in the portions embedded in the concrete, the diameter decreases linearly with distance, as shown in the following equation:

$$\emptyset(x) = \emptyset_0 + x \frac{(\alpha - 1)\emptyset_0}{L} \quad [6]$$

### Analytical Solution

In deriving the analytical solution for the DTR dowels, the original assumption of constant modulus of dowel support K is adopted, even though the modulus of dowel support is inversely related to the diameter of the dowel and to the depth of concrete below the dowel (Porter and Guinn 2002). All subsequent derivations correspond to a value of α = 0.20, meaning that the area of the double-tapered ends is only 0.04 times the central area of diameter $\emptyset_0$. We can express the relative stiffness β of the double tapered shaped dowel as function of x as follows:

$$\beta(x) = \sqrt[4]{\frac{16K}{\pi E_s \left[\emptyset_0 + x \frac{(\alpha - 1)\emptyset_0}{L}\right]^3}} \quad [7]$$

Substituting the expression for β(x) into the general equations for deflection y(x) (equation [1]) and vertical shear V(x) (equation [4]) respectively, we obtain the following equations:

$$y_t(x) = \frac{e^{-\beta(x)x}}{2\beta(x)^3 E_s I(x)}\left[P\cos[\beta(x)x] + \frac{\beta(x)Pz}{2}(\cos[\beta(x)x] - \sin[\beta(x)x])\right] \qquad [8]$$

$$V_t(x) = -e^{-\beta(x)x}[(-\beta(x)Pz - P)\sin[\beta(x)x] + P\cos[\beta(x)x]] \qquad [9]$$

At the face of the joint (x = 0), the deflection of the double-tapered dowel is:

$$y_t(0) = \frac{P}{2\beta_0^3 E_s I_0}\left[1 + \frac{\beta_0 z}{2}\right] \qquad [10]$$

Where

$$\beta_0 = \beta(0) = \sqrt[4]{\frac{16K}{\pi E_s \emptyset_0^3}}$$

$$I_0 = I(0) = \frac{\pi}{64}\emptyset_0^4$$

**Linear FE Analysis using Abaqus (2011)**

A FE analysis of the performance of DTR dowels as compared to traditional cylindrical dowels was performed using the FE code ABAQUS (2011). A total of five computational runs were performed, each with a varying parameter α that ranges from 0.0 to 1.0 in 0.2 increments. Again, all the simulated cases (as shown in the upper left table of Figure [3]) involve DTR dowels with the same weight of steel as a 35 mm traditional cylindrical dowel.

In the FE model, 16116 quadratic solid elements (C3D20R) were used to model the dowel bar, with an average element size of 2.5 mm. Due to symmetry, only half of the dowel was simulated. A typical embedment length of 250 mm was adopted. A Winkler foundation support was applied on the embedded surface of the dowel bar. The stiffness of the Winkler foundation is nothing but the modulus of dowel support K; a value of 150 N.mm$^{-3}$ was adopted. Symmetry boundary conditions were applied along the vertical mid-plane of the dowel.

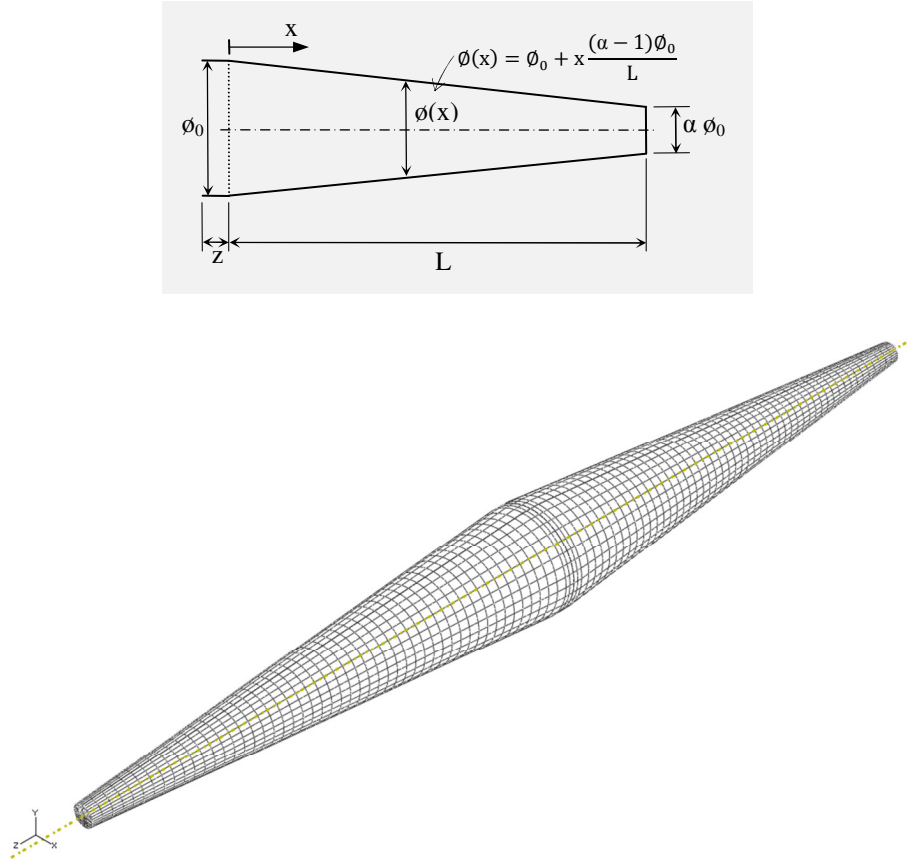

Figure 2 – (a) Geometry configuration of DTR dowel, and (b) 3D FE mesh of DTR dowel (full model)

The maximum shear force P at the midpoint of the dowel was applied as a distributed pressure over the circular cross section of the DTR dowel of diameter $\emptyset_0$. The value of P, the maximum shear force carried by the dowels is usually calculated by multiplying the magnitude of the applied wheel load by approximately 0.45 (Yoder & Witczak 1975) that accounts for the load transferred across the joint, then dividing the obtained number by the amount of effective dowels participating in load transfer. A value of P = 3000 N was chosen in order to later compare numerical results with analytical solutions. The moment $M_o$ applied at the face of the joint was calculated from the maximum shear P and the joint width z as $M_o = Pz/2$.

**Performance of Cylindrical and DTR Dowels**

In order to perform comparative analysis between traditional cylindrical dowels (of diameter ø and length L) and new dowel bar shapes, the same weight of steel material should be used for both dowel shapes. If the same weight of material can be used in DTR dowels to achieve better performance in terms of lower bearing stresses and higher amount of load transfer, then the dowel spacing can be increased. By equating the volume of the traditional dowel bar with the volume of the new tapered dowel, we obtain an expression relating the diameter ø of the traditional cylindrical dowel to the parameters $ø_0$ and α of the DTR dowel, as shown in the following equation:

$$\frac{ø}{ø_0} = \sqrt{\frac{z + \frac{2}{3}L(\frac{1-\alpha^3}{1-\alpha})}{2L + z}} \quad [11]$$

In the above equation, as α tends to unity, the above equation reduces to ø = $ø_0$, which corresponds to the case of traditional cylindrical dowels.

In order to compare the performance of traditional cylindrical dowels with DTR dowels, the following criteria were examined: (1) bearing stress on concrete and (2) vertical shear force along the dowel.

**Bearing Stress on Concrete (load-controlled)**

The bearing stress $\sigma_t(x)$ along the DTR dowel can be expressed by replacing the expression for β(x) shown in equation [7] into equation [1] for the deflection, and then multiplying y(x) by the modulus of dowel support K:

$$\sigma_t(x) = K\, y_t(x) = \frac{K\, e^{-\beta(x)x}}{2\beta(x)^3 E_s I(x)}\left[P\cos\beta(x)x + \frac{\beta(x)Pz}{2}(\cos\beta(x)x - \sin\beta(x)x)\right] \quad [12]$$

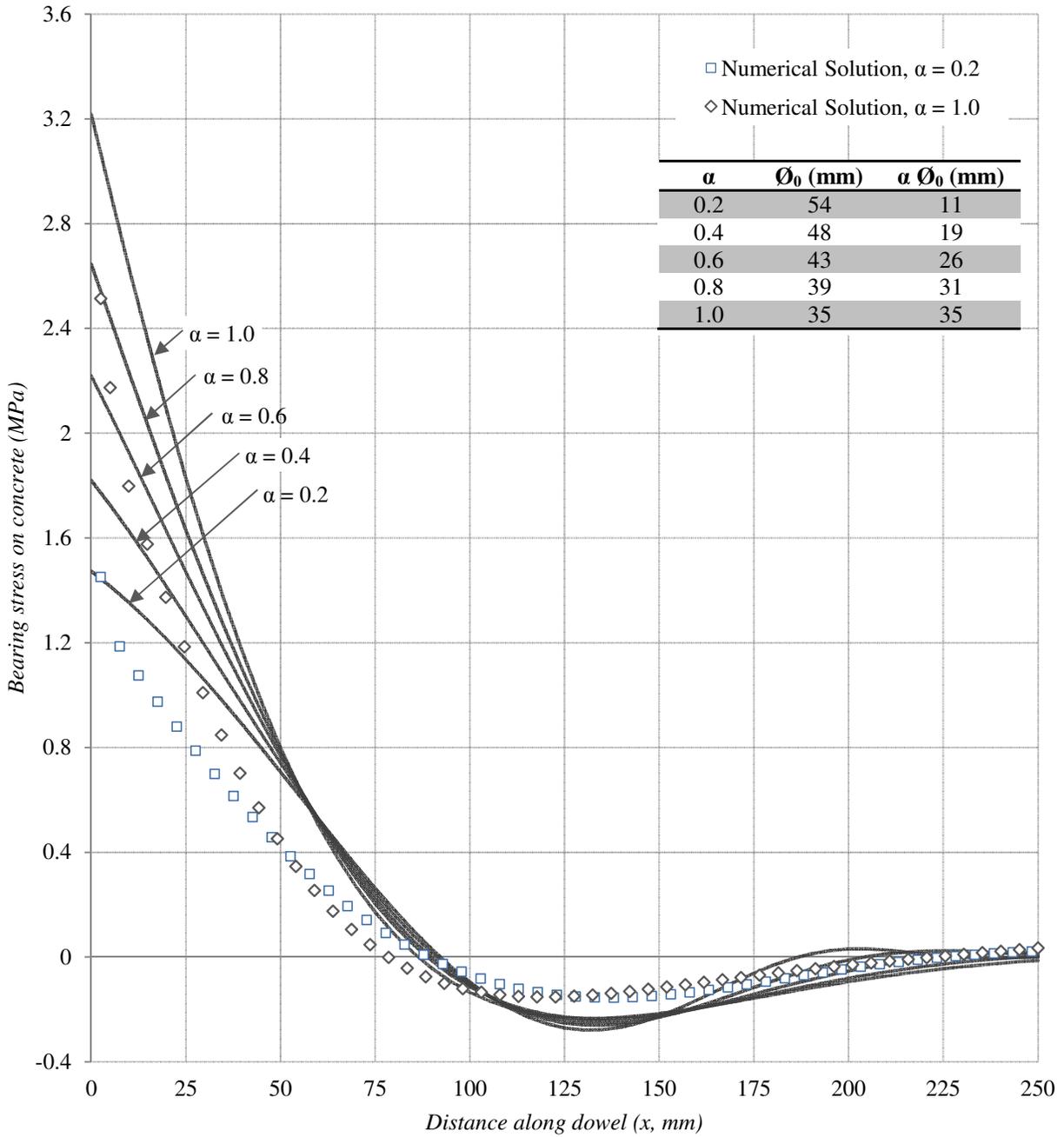

Figure 3 – Comparison of bearing stress along traditional cylindrical and DTR dowels of varying α, with $K = 150$ N.mm$^{-3}$, $z = 10$ mm, $P = 3000$ N, $E_S = 200000$ N.mm$^{-2}$, $L = 250$ mm (the same weight of material is used in all cases)

Figure [3] plots the analytical and numerical solutions for bearing stress along DTR dowels of equal weight but with varying geometry. In the analytical solution, equation [12] is plotted for several values of α (0.2, 0.4, 0.6, 0.8, and 1.0), as shown in the solid lines of Figure [3]. In the numerical solution, the extreme cases of α = 1.0 (cylindrical dowel) and α = 0.20 are plotted.

For the case of traditional cylindrical dowels (α = 1.0), Friberg's analytical solution, as compared to the numerical solution, overestimates the bearing stresses along the concrete, particularly near the joint faces, where the analytical solution is 16.1% higher than FE results. This relative discrepancy between the analytical and numerical solutions leads back to the conclusions of Mannava et al. (1999) that a uniform modulus of dowel support is insufficient for capturing the deformation profile (or bearing stress) along the dowel. The same discrepancy is observed between the analytical and numerical solutions for the case of α = 0.20. However, near the joint faces, the difference reduces to only 0.5%.

While keeping the same volume of material, as the slope of the tapered sides increases, the bearing stresses along the concrete drop, as shown in the solid lines of Figure [3]. This is particularly true near the joint face, where as the parameter α goes from 1.0 to 0.2, the bearing stress drop by approximately 2.2 times.

A summary of the cases investigated ranging from traditional cylindrical dowels (α = 1.0) to perfectly double-cone shaped dowel (α = 0.0) is shown in the table of figure [3]. Given the value of α that ranges from 0 to 1, the ratio ø/ø$_0$ was calculated using equation [11] for ø = 35 mm in order to compare the performance of dowels for the same weight of material.

In order to compare the performance of the proposed DTR dowel with traditional cylindrical dowels, (1) the bearing stress for the DTR dowel $\sigma_{t0}$ at the face of the joint (x = 0.0) is normalized to the corresponding bearing stress $\sigma_0$ for a typical ø = 35 mm cylindrical dowel. We recall that ø$_0$ is the maximum central diameter of the DTR dowel. The bearing stress ratio $\sigma_{t0}/\sigma_0$ can be expressed using the expression for the deflection of the embedded portion of the dowel at the joint face for cylindrical (y$_0$) and DTR dowels (y$_{t0}$):

$$\frac{\sigma_{t,0}}{\sigma_0} = \frac{y_{t,0}}{y_0} = \frac{\beta^3 I \, (2 + \beta_0 z)}{\beta_0^3 I_0 \, (2 + \beta z)} \qquad [13]$$

Since we are dealing with relatively small joint widths, we can further assume that the ratios (2+β$_0$z) ≈ (2+βz) ≈ 2, hence reducing the above equation into:

$$\frac{\sigma_{t,0}}{\sigma_0} = \left(\frac{\phi_0}{\phi}\right)^{-1.75} \qquad [14]$$

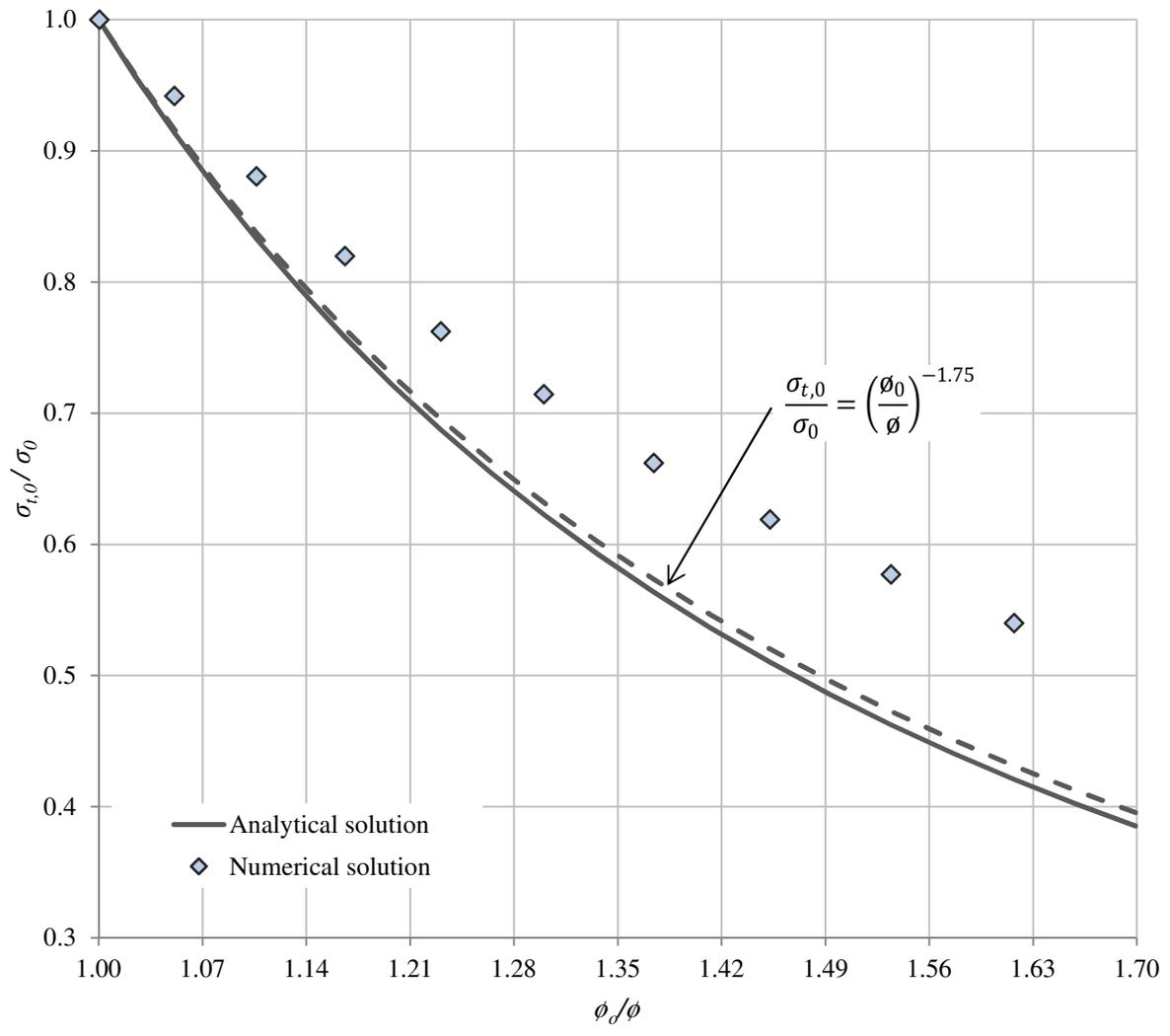

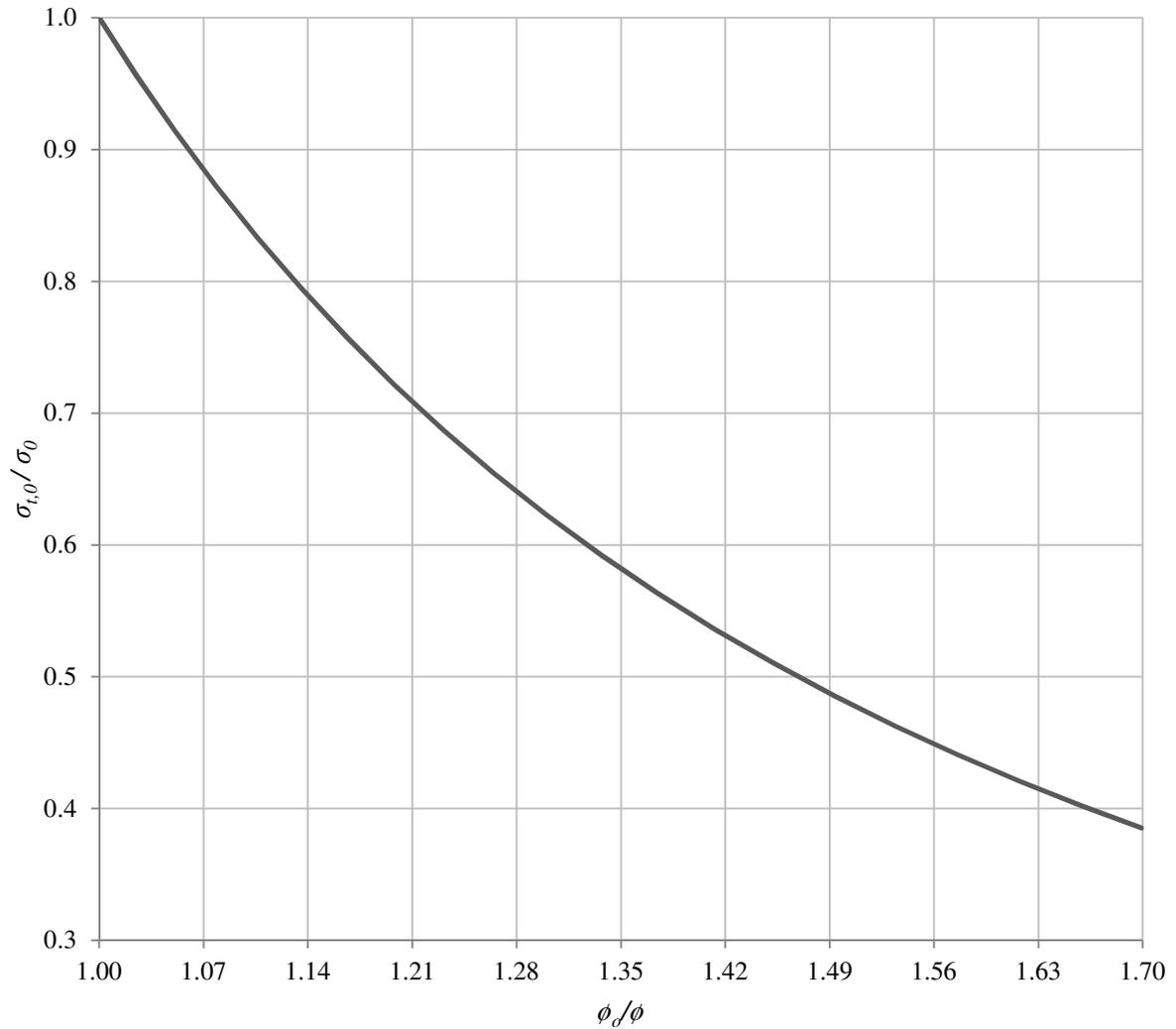

Figure 4 – Normalized bearing stress as a function of dowel parameter α (or ø/ø$_0$)

Figure [4] plots the normalized bearing stress ($\sigma_{t0}/\sigma_0$) as a function of the diameter ratio ø$_0$/ø. We recall that $\sigma_{t0}$ and $\sigma_0$ represent the bearing stresses at the face of the joint for the DTR and cylindrical dowels, respectively. While the analytical solution matches the numerical solution at ø$_0$/ø = 1.0 (or α = 1, cylindrical dowel), as the ratio ø$_0$/ø increases (the slopes of the dowel become more and more tapered, for the same weight of material), the analytical solution diverges from the numerical solution. For the extreme case of α = 0.2 (ø$_0$/ø = 1.5), the analytical solution differs from the numerical solution by approximately 26%, and the difference decreases as α increases to 1.0. One of the main reasons of divergence between the analytical and numerical

solutions is that the modulus of dowel support, as mentioned earlier, is dependent on the dowel diameter. Thus, the assumption of uniform modulus of dowel support along the DTR dowel (with varying diameter) should be verified. Since the modulus of dowel support is inversely related to the diameter of the dowel, in the vicinity of the joint face, where the diameter of the DTR dowel is highest, the modulus of dowel support K should be appropriately reduced.

In addition, equation [14], which is an approximation of the analytical solution, is plotted on Figure [4]. Equation [14] slightly differs from the analytical solution, mainly because of the assumption of negligible joint width z.

The advantages of using DTR as compared to cylindrical dowels in terms of reduced bearing stresses are visualized in Figure [4]. As the slopes of the tapered sides of the DTR dowel increase (the ratio $ø_0/ø$ increase), the bearing stress ratio $σ_{t0}/σ_0$ at the face of the joint drops. According to the numerical solution for the case of $α = 0.20$, (or $ø_0/ø = 1.5$), the bearing stress at the joint face for the DTR dowel is only 58% of the corresponding bearing stress for the traditional cylindrical dowel.

**Vertical Shear Force in Dowel (displacement-controlled)**

In order to compare the load-transferring capacity of both traditional cylindrical and DTR dowels, the amount of shear force corresponding to a specific amount of deflection in the dowel is determined. In other words, a vertical downward displacement is imposed on the dowel at the joint face and the corresponding shear force is calculated. Using the deflection equation of the dowel at the face of the joint, the maximum shear force P corresponding to $y_0 = 0.01$ mm is computed for all other parameters equal ($E_{Steel} = 200,000$ MPa, $K = 150$ N.mm$^{-3}$, $L = 250$ mm, $ø = 35$ mm, and $z = 10$ mm), as shown in the following sample calculation. For the cylindrical dowel ($α = 1.0$), the maximum shear force $P_0$ necessary to induce a 0.01 mm vertical deflection at $x = 0.0$ is:

$$P = \frac{2β^3 E_s I}{[1 + βz/2]} y_0$$

Solving for $P = 1400$ N. For the DTR dowel with $α = 0.20$, the maximum shear force $P_t$ necessary to induce a 0.01 mm vertical deflection at $x = 0.0$ is:

$$P_t = \frac{2\beta_0{}^3 E_s I_0}{[1 + \beta_0 z/2]} y_{t,0}$$

Solving for $P_t$ = 3024 N.

The distribution of shear along the dowel is then plotted using equation [10]. We recall that for cylindrical dowels ($\alpha$ = 1.0), equation [10] reduces to the traditional equation of shear along the dowel presented by Friberg (1938), with $\beta(x)$ given by equation [7]. The shear distribution is plotted in Figure [5] with values of $\alpha$ ranging from 0.2 to 1.0 in 0.2 increments. The corresponding central diameters $\emptyset_0$ for each DTR dowel are shown in the table of Figure [5]. At this point, we recall two things, (1) $\alpha\emptyset_0$ represents the end diameter of the DTR dowel, and (2) all dowels shown with $\alpha$ ranging from 0.2 to 1.0 have the same weight.

The use of DTR dowels, particularly for low values of $\alpha$, increases the amount of load transferred through vertical shear, as shown in Figure [5]. As the value of $\alpha$ drops from 1.0 (which represents cylindrical dowel) to 0.2, the corresponding maximum shear force P at the joint face increases. This implies that the amount of load transferred via the dowel is directly proportional to the slope of the DTR dowel. The highest load-transfer is achieved for the DTR dowel with the lowest $\alpha$. In fact, as $\alpha$ drop from 1.0 to 0.2, the maximum shear force along the dowel increases by 116%. The fact that the DTR dowel has a better load-transfer capability than traditional cylindrical dowel can be significantly beneficial in extending the life of doweled rigid pavements.

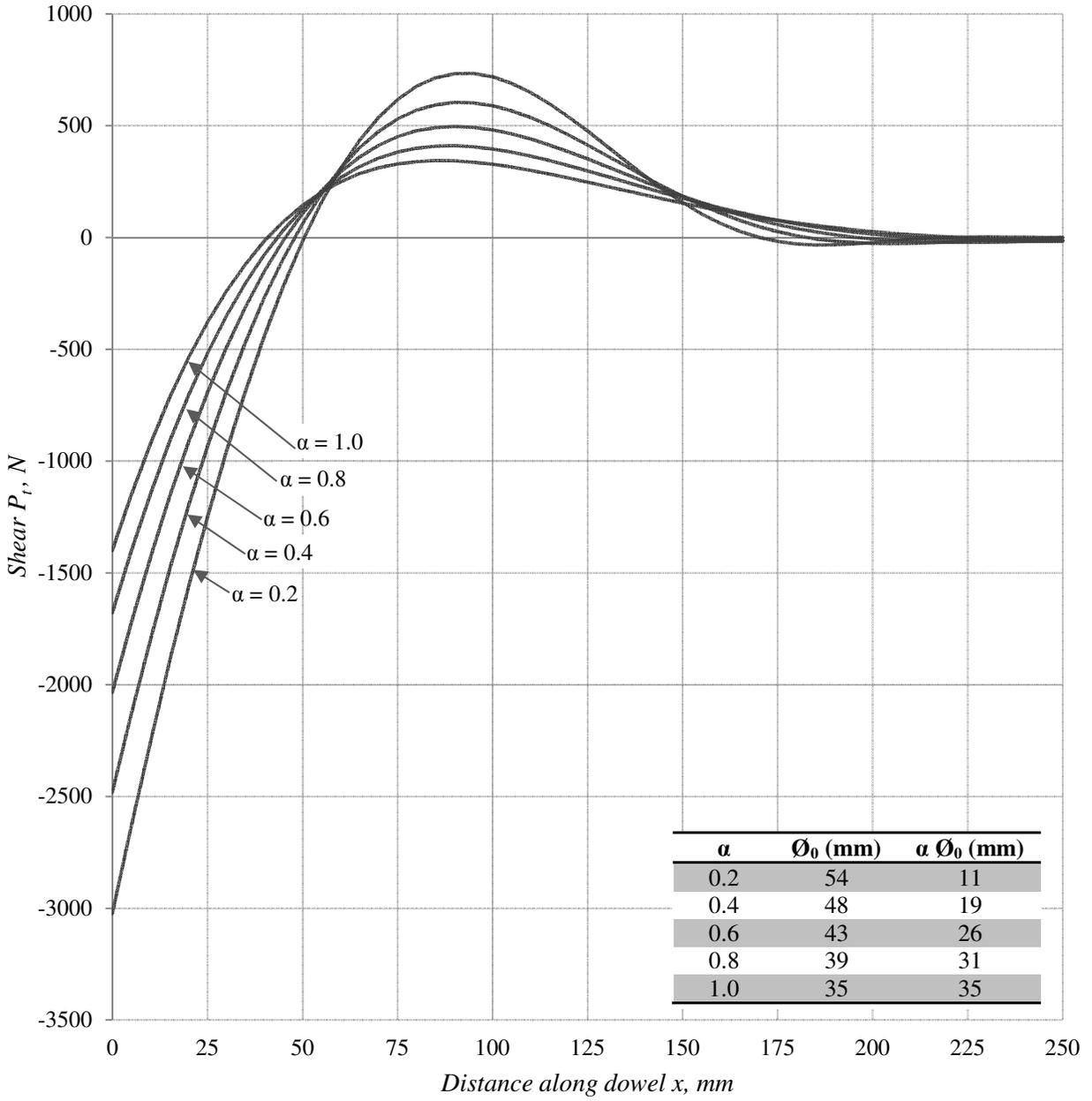

Figure 5 - Comparison of shear force along cylindrical and DTR dowels corresponding to $y_0 = 0.01$ mm

**Non-Linear FE Analysis using Abaqus (2011)**

In order to further compare the performance of traditional cylindrical and DTR dowels, a non-linear FE Analysis was conducted using Abaqus (2011). The model includes a steel dowel embedded in a concrete mass. The dowel extends from the joint face by a distance of z/2, where z is the joint width. Due to symmetry, only half of the assembly was modeled, as shown in Figure [6], and z-symmetry was imposed along the vertical x-y mid-plane. Fixed BC's were imposed along the other boundaries, except the vertical joint face. A biased mesh was implemented, as shown in Figure [6], with the smallest elements in the vicinity of the dowel. Quadratic solid elements (C3D20R) were used to model both the dowel and the surrounding concrete. Full bonding between the dowel and the concrete matrix was assumed.

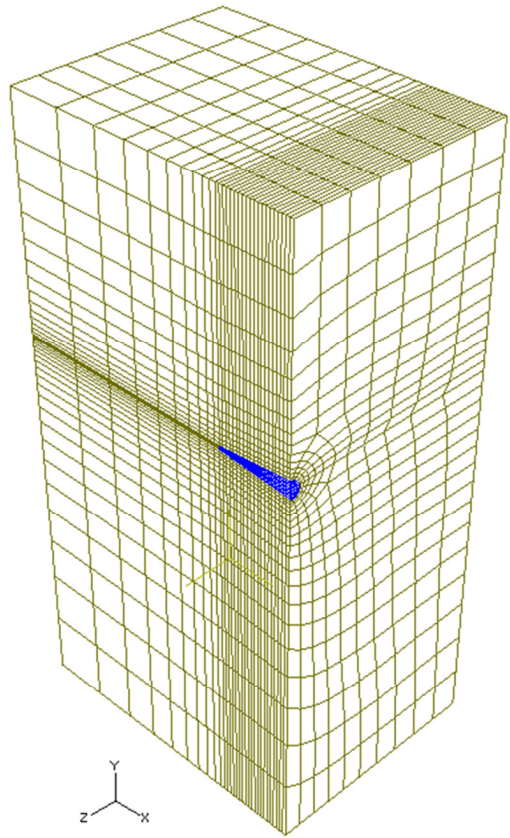

Figure 6 – 3D Finite Element mesh configuration of non-linear model.

Table 2 – Elastic-Plastic properties for steel material and parameters for the Concrete Damaged Plasticity (CDP) model

| Steel Properties | |
|---|---|
| Young's modulus (MPa) | 200000 |
| Poisson's ratio | 0.3 |
| Yield stress (MPa) | 345 |
| **Concrete Properties** | |
| Young's modulus (MPa) | 31027 |
| Poisson's ratio | 0.15 |
| Dilation angle | 36.31° |
| Compressive initial yield stress (MPa) | 13 |
| Compressive ultimate stress (MPa) | 24.1 |
| Tensile failure stress (MPa) | 2.9 |

An Elastic-Plastic model for the steel material in the dowel bar was adopted, as shown in Table [2]. The concrete mass was modeled in Abaqus/Standard using the Concrete Damaged Plasticity (CDP) model (Lubliner et al. 1989, Lee and Fenves 1998). In the CDP model, isotropic damaged elasticity is coupled with isotropic compressive and tensile plasticity. In tension, softening response occurs and post-failure behavior is specified in terms of a stress-displacement curve, while in compression, initial hardening is followed by softening behavior. The parameters of the concrete damaged plasticity model were obtained from Bhattacharjee et al. (1993) & Ghrib and Tinawi (1995), as shown in Table [2].

The solution was obtained using the Riks method in Abaqus/Standard, given that significant nonlinearity is expected in the solution as the concrete degrades and the dowel yields. An automatic incrementation was adopted, with an initial increment size of 0.001. An ultimate load of 150 KN was applied as shear traction over the free end of the dowel, for a total time period of 1.0. Since only half of the assembly was modeled, half of the ultimate load was applied at the dowel free end.

**Stress-Displacement Response**

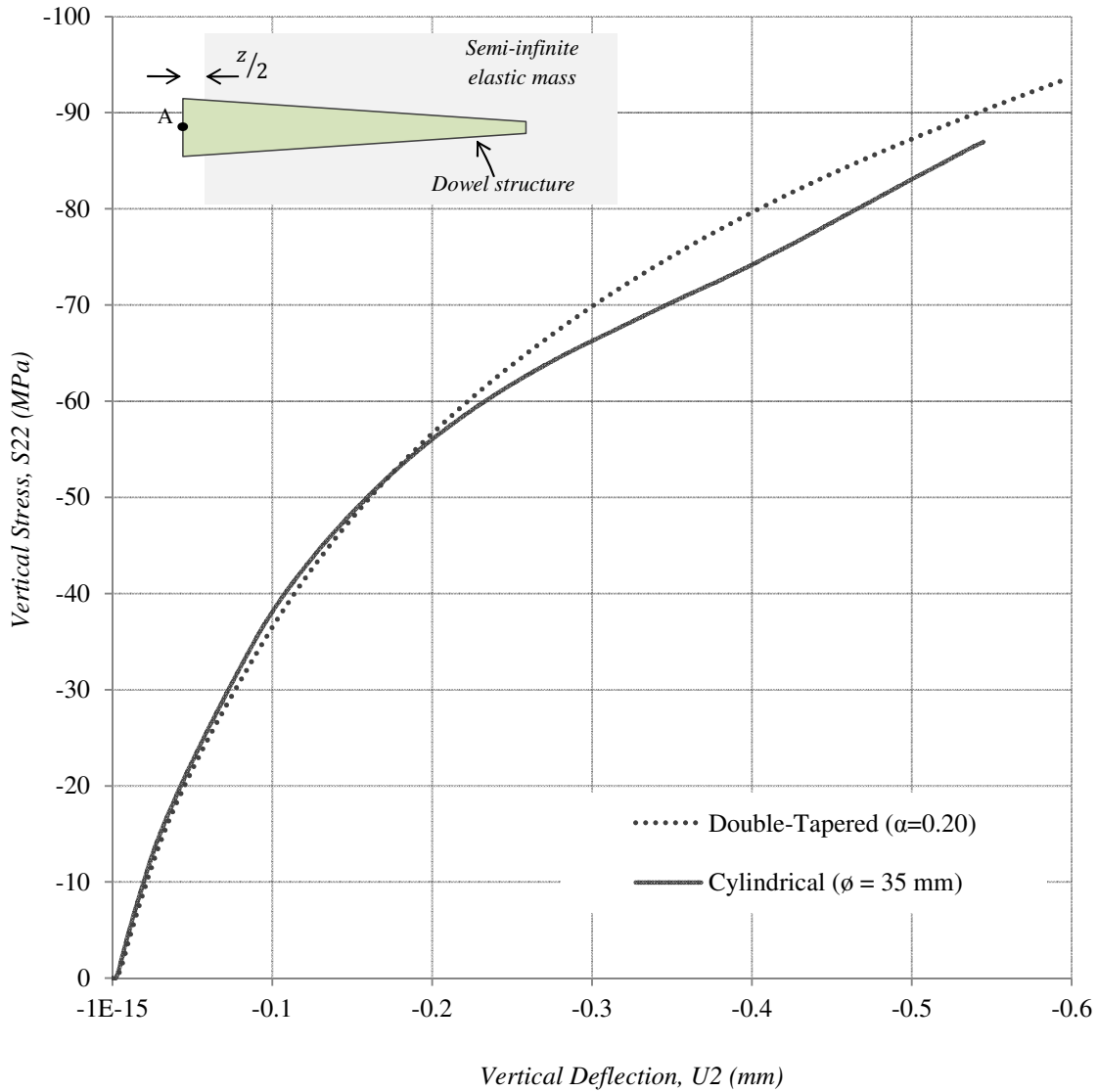

Figure 7 – Vertical stress versus deflection at the midpoint (point A) of DTR ($\alpha = 0.20$) and cylindrical dowels ($\varnothing = 35$ mm)

Figure [7] visualizes the stress-displacement response at the midpoint of both DTR and cylindrical dowel. The midpoint of the dowel is represented by point A in the upper left diagram on Figure [7]. It can be observed that DTR dowels have mainly two advantages over cylindrical dowels in the inelastic range: (1) higher load-carrying capacity, particularly as the concrete surrounding the dowel degrades and the steel in the dowel itself yields; (2) lower vertical

differential deflections across the concrete floors. The above two point are extremely beneficial in extending the service life of the doweled rigid pavement under repetitive loading.

**Damage Propagation Analysis**

The stiffness degradation (SDEG) of the concrete matrix around the dowel is plotted in Figure [8a and 8b] in the longitudinal (x/L) and vertical directions (y/ø or y/ø$_0$) respectively. SDEG profiles for both the DTR and cylindrical dowels were obtained at two steps corresponding to a downward vertical displacement at the dowel midpoint (point A) of 0.1 mm and 0.5 mm. We bear in mind that according to Friberg's (1938) analytical model, a displacement at the dowel midpoint of 0.5 mm corresponds to a total differential deflection across slabs of 1 mm.

The stiffness damage scalar is plotted in the longitudinal direction (x / L) in Figure [8a]. It can be noted that the damage in the concrete was confined to the face of the joint; above x/L = 0.35, practically no damage occurs.

The level of SDEG in the concrete with the DTR dowel was higher than that of the case of the traditional cylindrical dowel, for both midpoint deflections of -0.1 and -0.5 mm. This implies that the use of DTR dowel, as compared to cylindrical dowels, is beneficial in terms of reduced differential deflection across slabs. In fact, in order to achieve the same level of differential deflection (=twice the deflection at point A), the DTR dowel allows for more concrete degradation than the cylindrical dowel. The cylindrical dowel deflects by -0.1 mm when the SDEG parameter at the face of the joint is equal to 0.48, while for the DTR dowel, the same deflection is reached when degradation of the concrete at the joint face reaches 0.65. The same is true for the case of U2 @ A = -0.5 mm, where the DTR dowel allows for higher concrete degradation as compared to the cylindrical dowel before if deflects by 0.5 mm at its midpoint (point A).

This leads to the idea that the performance of DTR dowels is better than the cylindrical dowel not only in the elastic range, but even after the concrete matrix surrounding the dowel has degraded. In fact, it can be concluded from Figure [8a] that with the degradation of concrete around dowels under repetitive loading, the differential deflection across slabs with DTR dowels is significantly lower than the corresponding value with cylindrical dowels.

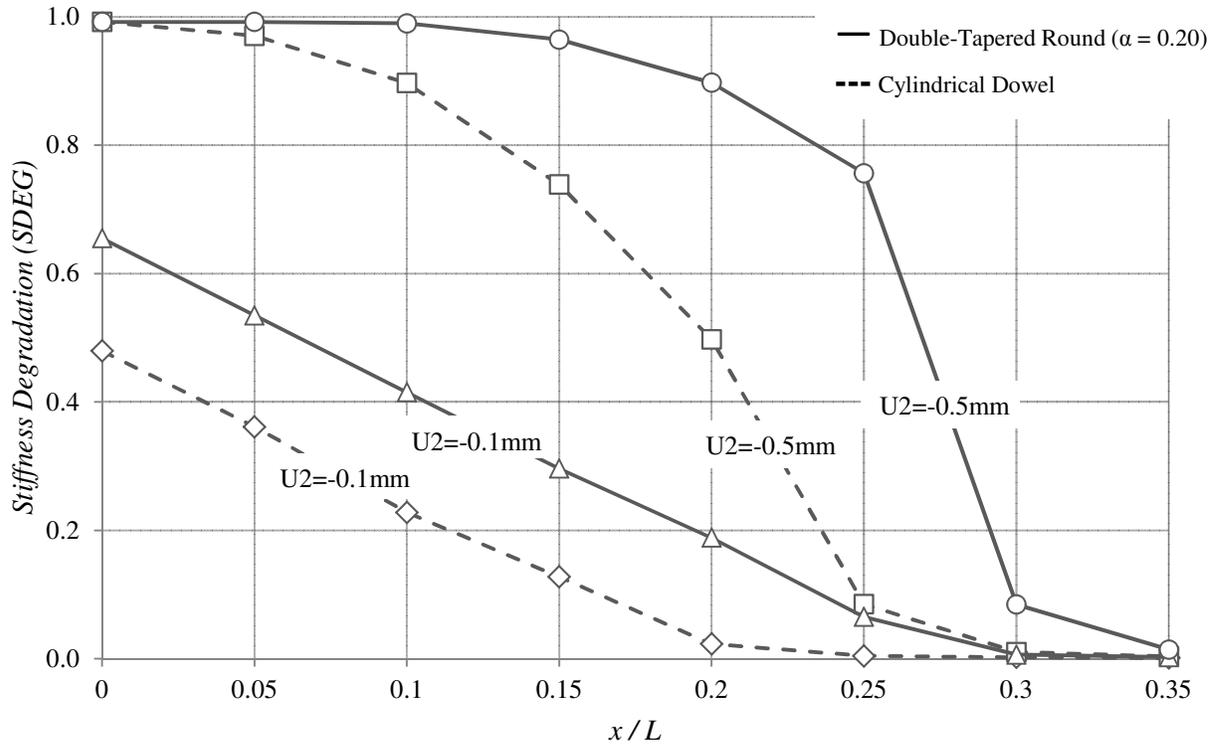

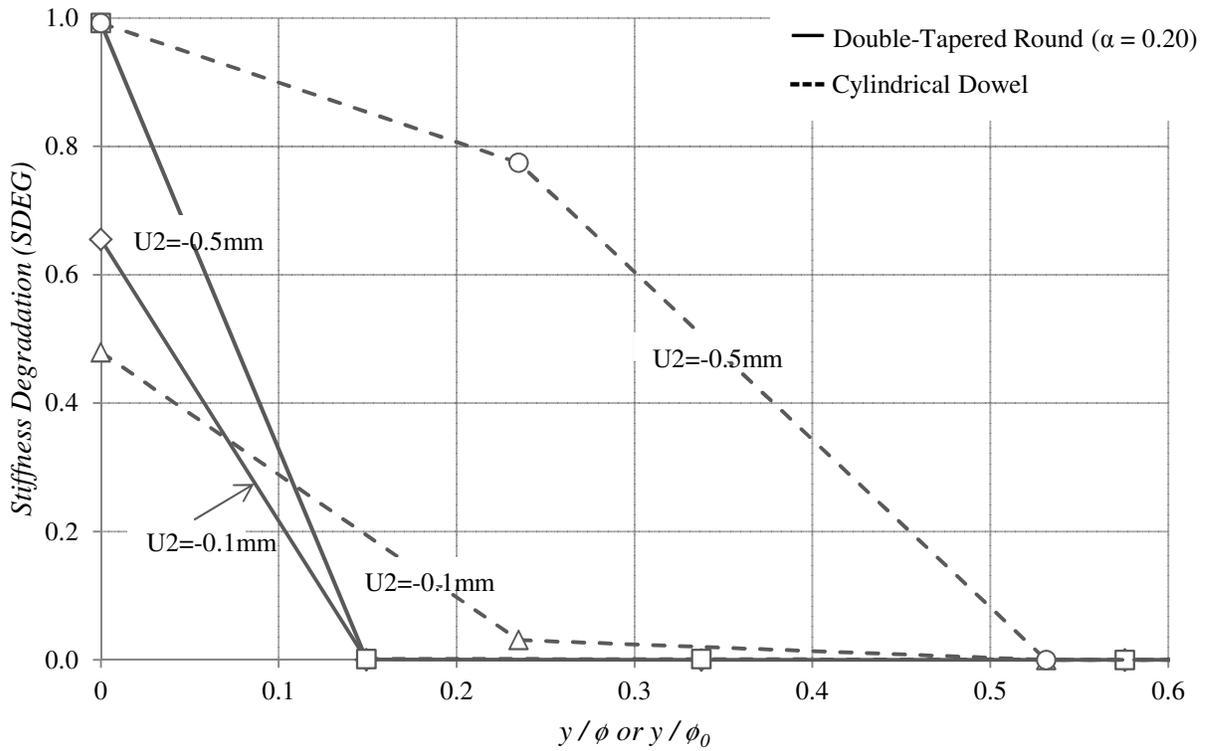

Figure 8 - Stiffness degradation in concrete matrix for DTR and cylindrical dowels (a) along x/L ($0 \leq x/L \leq 0.35$), and (b) along $y/\phi$ or $y/\phi_0$

Figure [8b] plots the stiffness degradation scalar in the concrete at the face of the joint in the vertical direction; y is a downward vertical axis, whose plane is the face of the joint, and with origin the bottom of the dowel (@ x = 0.0). The value of y is normalized to the diameter $ø_0$ in the case of DTR dowels, and ø in the case of cylindrical dowels. Results are shown at two midpoint dowel deflections of -0.1 and -0.5 mm.

It can be observed first from Figure [8b] that for the case of the traditional cylindrical dowel, stiffness degradation was confined within a distance of y/ø = 0.53, i.e. damage propagated downward by 0.53 ø, whereas for the case of the DTR dowel, stiffness degradation occurred within a distance of $y/ø_0$ = 0.15, or damage propagated downward by $0.15ø_0$. This implies that one of the advantages of using DTR dowels over cylindrical dowels is in confining the damage in the concrete in the vicinity of the dowel. As shown in Figure [8b], damage propagation below the DTR dowel is only approximately 28% of the corresponding damage below the cylindrical dowel.

## Conclusions and Recommendations

This study numerically investigated the performance of traditional round dowels in concrete floors. The design of traditional cylindrical dowels was optimized through Finite Element (FE) analysis. A new type of Double-Tapered Round (DTR) dowels was proposed, and the latter's superior performance over traditional cylindrical dowels was verified through LE and nonlinear FE analysis in Abaqus (v-6.11). The study found that for the same weight of steel material in the dowel, the bearing stress at the face of the joint dropped by approximately 2.2 times as α goes from 1.0 to 0.2 and the slope of the tapered sides increased.

The analytical and numerical solution for the DTR dowel slightly diverged as α dropped from 1.0 to 0.2 due to varying modulus of dowel support K. While K is inversely related to the diameter of the dowel, near the joint face, the diameter of the DTR dowel is highest, and K should be adequately reduced.

As for the amount of vertical load-transfer through shear, DTR dowels were found more effective in transferring the load across concrete slabs. As the dowel slopes become more and more tapered (α goes from 1.0 to 0.2), the load-transfer capability of the dowel increased by as far as 116%. This underlines the fact that DTR dowels have a better load-transfer capacity than traditional cylindrical dowel, something that can extend the life of the doweled rigid pavement under cyclic loading.

In the inelastic range, the performance of DTR dowels outweighed that of traditional cylindrical dowels. With the degradation of the concrete matrix and steel yielding, the DTR dowel maintained a higher load-transfer capacity than the traditional cylindrical dowel, and presented lower amounts of differential deflections across concrete floors. In fact, even after extensive degradation of concrete around the dowels, the differential deflection across slabs with DTR dowels was significantly lower than the corresponding value with cylindrical dowels. In order to reach the same level of differential deflection, the DTR dowel allows for more concrete degradation than the cylindrical dowel.

For both DTR and cylindrical dowels, damage was practically confined at the face of the joint; beyond a distance of x / L of 0.35, no degradation of concrete matrix occurred. In the vertical direction at the face of the joint, for cylindrical traditional dowels, damage occurred within a distance of 0.53 ø below the dowel, while for the DTR dowel, damage propagated a distance of only 0.15 $ø_0$ below the dowel. This leads to the idea that DTR dowel are more effective in confining the damage to the vicinity of the dowel.

Several recommendations on the design of DTR dowels can be presented; among other things, for ease of manufacturing, the diameter of the DTR dowel can continue to linearly increase along the joint. The results presented earlier would not be affected whether the dowel diameter is constant ($ø_0$) or linearly tapered along the joint, mainly because the joint has negligible width compared to the DTR dowel dimensions.

In order for the DTR dowel to function properly and present sufficient load transfer, the bottom and the top of the dowel must be in complete contact with the surrounding concrete. Void space should be left on the lateral sides of the DTR dowel through the sleeve assembly, in order to allow for unrestrained lateral displacement. Similarly, void space should be left at the end of the dowel to allow for thermal expansion/contraction. In order to achieve optimum design of DTR dowels, the value of α should be chosen as small as possible, yet not too small in order to allow for a void space near the embedded end of the dowel.

One of the potential difficulties that would be encountered in using DTR dowels would be in the design of dowel sleeves in order to (1) provide a void space in the horizontal direction to allow for expansion of the dowel under increased temperatures, and (2) account for dowel misalignment. In the vertical direction, the sleeve should be in contact with the tapered-dowel faces in order to allow for a fully efficient load transfer between adjacent concrete slabs.

**List of Abbreviations**

$y$ = vertical deflection along the cylindrical dowel (L)

$y_t$ = vertical deflection along the double-tapered dowel (L)

$\sigma$ = bearing stress on concrete for cylindrical dowel (FL$^{-2}$)

$\sigma_t$ = bearing stress on concrete for double-tapered round dowel (FL$^{-2}$)

$E_S$ = modulus of elasticity of steel (FL$^{-2}$)

$P$ = maximum shear force carried by the dowel along the joint (F)

$I$ = moment of inertia of dowel (L$^4$)

$z$ = joint width (L)

$\beta$ = relative stiffness of dowel (L$^{-1}$)

$K$ = modulus of dowel support (FL$^{-3}$)

$\emptyset$ = diameter of cylindrical round dowel (L)

$V$ = vertical shear force along the cylindrical dowel (F)

$V_t$ = vertical shear force along the double-tapered dowel (F)

$\emptyset_0$ = central diameter of double-tapered round dowel (L)

$\alpha$ = ratio of double-tapered dowel end diameter to central diameter $\emptyset_0$

L = embedment length of dowel (L)

$f_c'$ = compressive strength of concrete (FL$^{-2}$)

D.T.R. dowel = double-tapered round dowel